\documentclass[twocolumn,prx,aps,longbibliography]{revtex4-1} 

\usepackage{color} 
\usepackage{natbib}
\usepackage{amsmath} 
\usepackage{multirow}  
\usepackage{graphicx} 
\usepackage{soul}
\usepackage{url}
\usepackage{textcase}
\usepackage{bm}
\usepackage{epstopdf}
\usepackage{enumitem}

\newlength{\figurewidth} 
\ifdim\columnwidth<20cm 
\else 
\fi 
\definecolor{burntorange}{RGB}{191,87,0}
\definecolor{maskgray}{RGB}{200,200,200}

\begin{document} 

\title{Evolutionary Dynamics of Cultural Memes and Application to Massive Movie Data}
\author{Seungkyu Shin}
\affiliation{Graduate School of Culture Technology and BK21 Plus Postgraduate Programme for Content Science, Korea Advanced Institute of Science \& Technology, Daejeon, Republic of Korea 34141}
\author{Juyong Park}
\affiliation{Graduate School of Culture Technology and BK21 Plus Postgraduate Programme for Content Science, Korea Advanced Institute of Science \& Technology, Daejeon, Republic of Korea 34141}
\affiliation{Sainsbury Laboratory, University of Cambridge, Cambridge, United Kingdom CB2 1LR}

\begin{abstract}
The profound impact of Darwin's theory of evolution on biology has led to the acceptance of the theory in many complex systems that lie well beyond its original domain. Culture is one example that also exhibits key Darwinian evolutionary properties: Differential adoption of cultural variants (variation and selection), new entities imitating older ones (inheritance), and convergence toward the most suitable state (adaptation).  In this work we present a framework for capturing the details of the evolutionary dynamics in cultural systems on the ``meme''---the cultural analog of the biological gene---level, and analyze large-scale, comprehensive movie--meme association data to construct a timeline of the history of cinema via the evolution of genres and the rise and fall of prominent sub-genres. We also identify the impactful movies that were harbingers to popular memes that we may say correspond to the proverbial ``Eve'' of the human race, shining light on the process by which certain genres form and grow. Finally, we measure how the impact of movies correlates with the experts' and the public's assessment.
\end{abstract} 

\maketitle 

\section{Introduction} 
Charles Darwin's theory of evolution has been highly influential in understanding the universal properties of systems that reproduce and change over generations.  In his seminal work \emph{The Origin of Species}, Darwin postulated that all organisms share a common ancestor, change over time through natural selection, and give rise to new species that underlie the impressive present-day diversity of living things on Earth. On the most fundamental level, Darwinian evolution stipulates that organisms evolve to adapt to the environment by an iterative process comprising variation, competition, and inheritance~\cite{darwin2009origin,fisher1999genetical,schwartz1978origins}. Its profound contribution to the understanding of biological evolution has prompted an active effort to apply the theory to many domains other than biology, leading to the coinage of the expression ``Universal Darwinism''~\cite{dawkins2006selfish}. Culture is one example domain where evolution with identifiable key Darwinian properties can be observed, evolving through a differential adoption of cultural variants in a manner analogous to the evolution of biological species~\cite{benedict2005patterns,cavalli1981cultural,cavalli1982theory,diamond1998guns,durham2001cultural,mesoudi2004perspective}: A new cultural product displays detectable variations from the others, competes with others in the marketplace to be selected by consumers. Successful variants are selected and thrive over others that may then perish and disappear. The successful variants further inspire the creators of later products to imitate or inherit their properties and further adapt~\cite{guglielmino1995cultural,hallpike1986principles}. Cultural evolution is thus the idea that beliefs, knowledge, artifacts, and human creations that constitute culture undergo a deeply analogous process by which species evolve through selective retention of favorable variants. In fact, in \emph{The Origin of Speies}, Darwin himself frequently cited cultural changes (primarily linguistic developments) to illustrate his theory of biological evolution~\cite{darwin2009origin}.

\begin{figure*} 
\includegraphics[width=160mm]{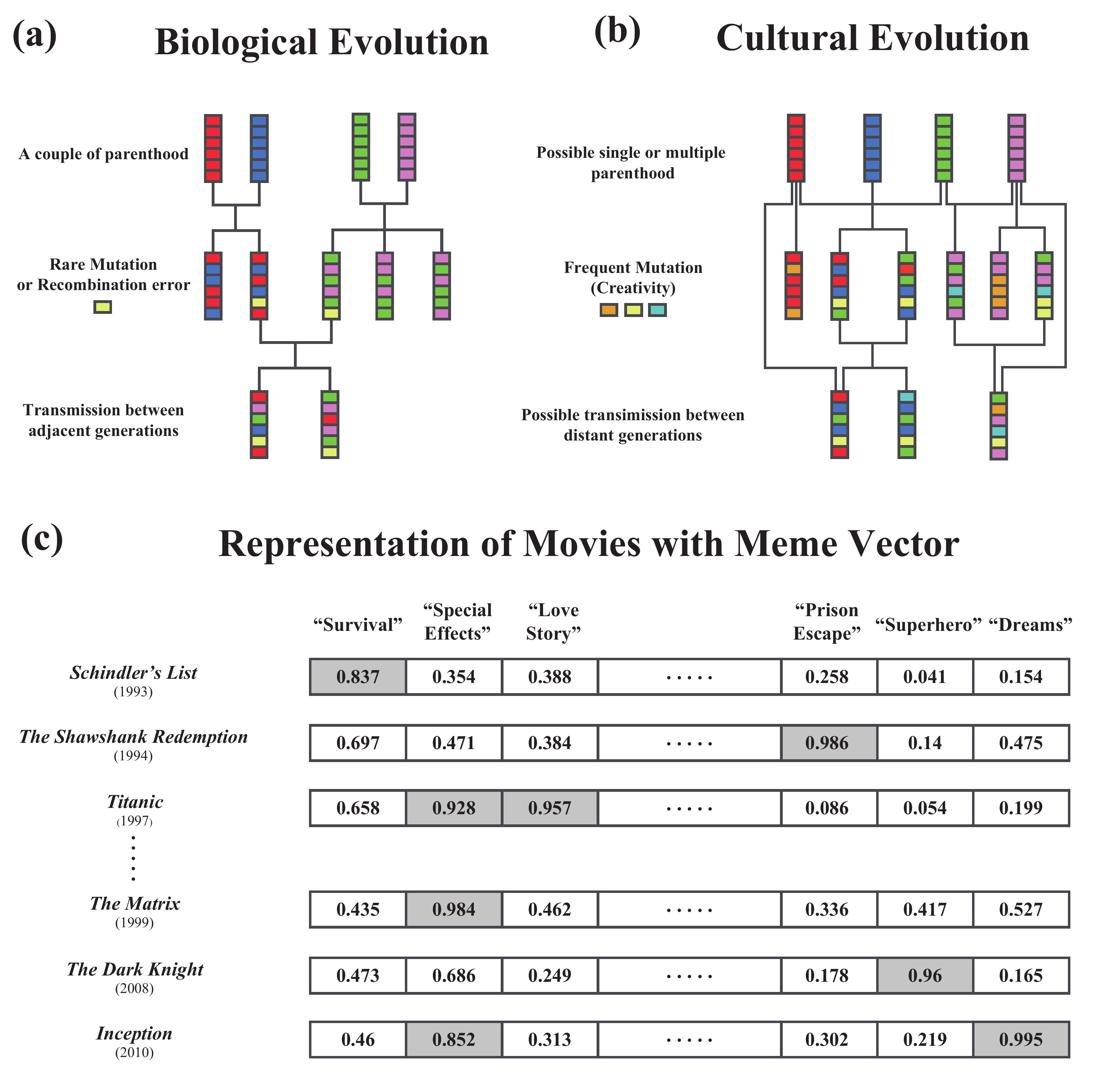} 
\caption{The analogy and the differences between evolutions in (a) biology on the gene level, and (b) culture on the meme level. Cultural evolution features an arbitrary number (from single to multiple) of parents, frequent mutations, and transmission between distant generations of memes. (c) Cultural products (movies in our paper) can be represented as a meme vector of the relevance score (association strength) with each meme. We use a meme vector of dimension 900, providing a rich set of descriptors for movies. The shaded components indicate the memes with particularly high relevance score ($0.8$ or greater).}
\label{figure01} 
\end{figure*} 

\begin{figure*} 
\includegraphics[width=160mm]{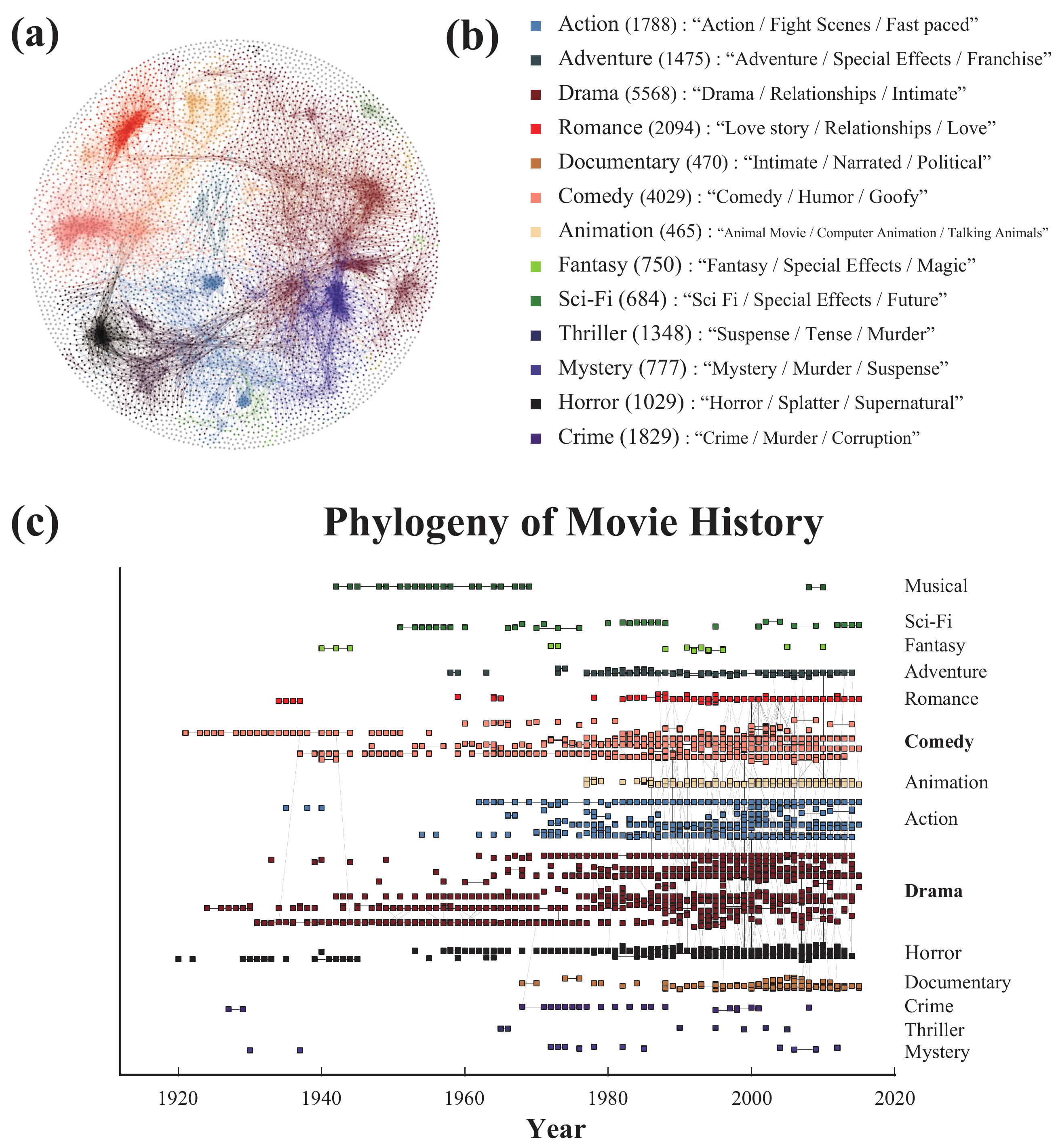} 
\caption{The correspondence between the network communities of movies and genres. (a) Network of movies based on the meme vector similarity between movies (PCC larger than $0.8$).  The communities detected in the network (colored) correspond well to established genres.  (b) Top three relevant memes within each genre. (c) Phylogeny of movie history showing how the movie genres have grown over time. The movie production years are given in the $x$-axis. Cross-linkage between genres is shown to increase over time, leading to hybrid genres such as romantic comedy (romance and comedy).} 
\label{figure02} 
\end{figure*}

\begin{figure*} 
\includegraphics[width=160mm]{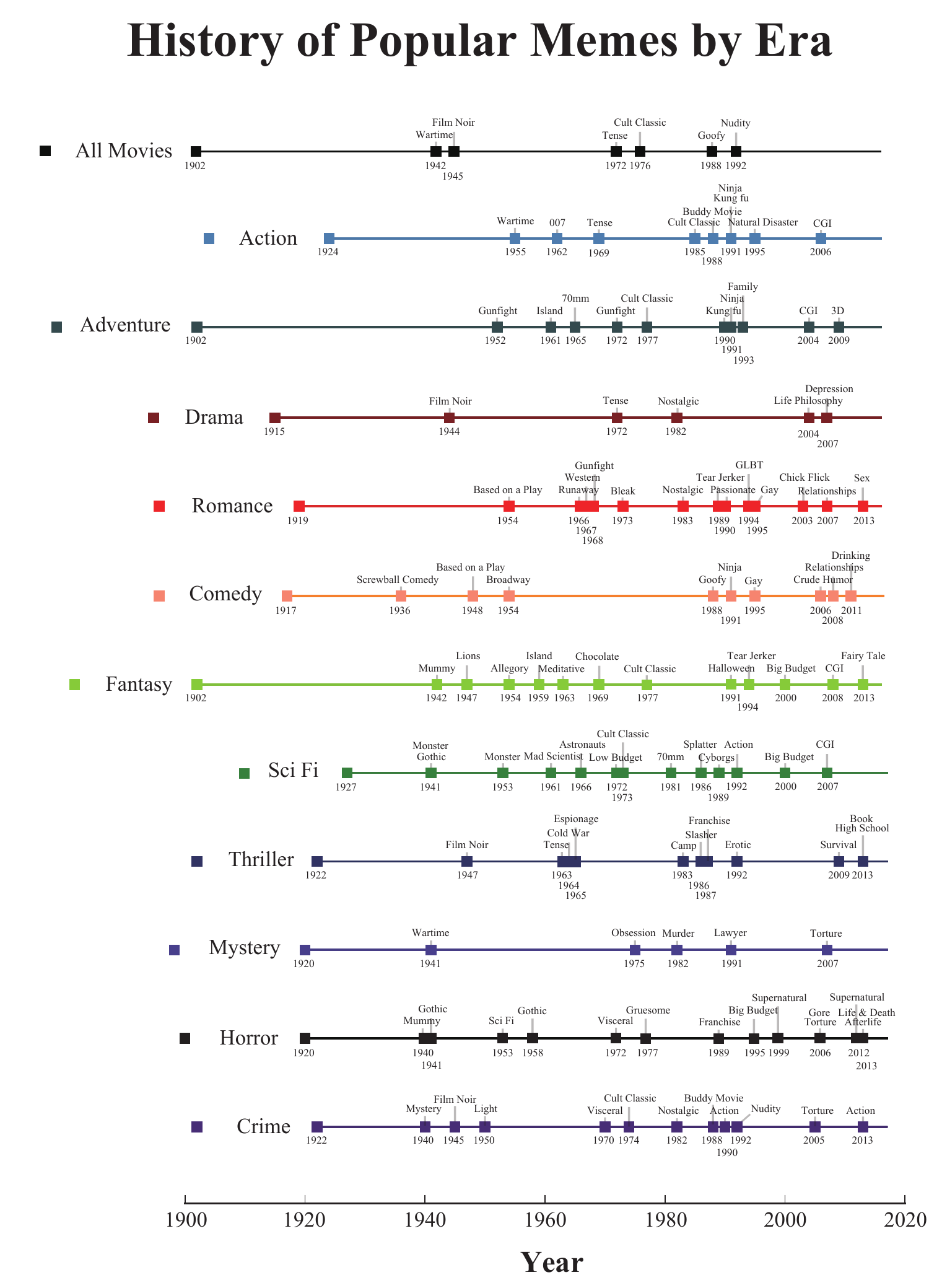} 
\caption{History of popular memes by era. The top row shows the most popular memes over all the movies, while the rest show those within each genre.} 
\label{figure03} 
\end{figure*}

\begin{figure*} 
\includegraphics[width=140mm]{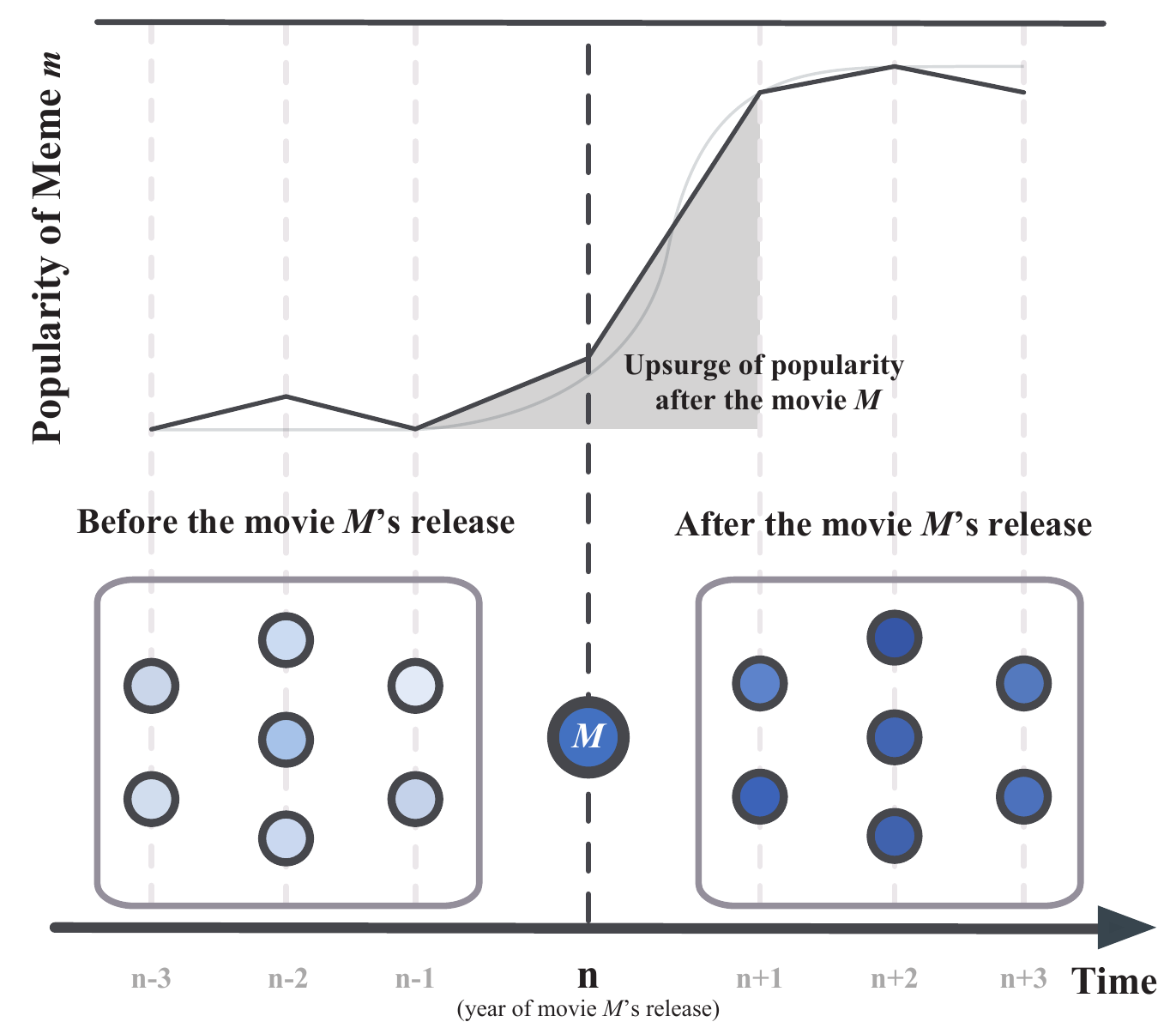} 
\caption{Defining the impact of a movie on the rise in the popularity of a meme. Such a movie features two characteristics, it precedes the meme's rise (top), and it has be highly relevant to the meme (bottom).}
\label{figure04} 
\end{figure*}

To make a deeper analogy with the current understanding of biological evolution, we study cultural evolution occurring on the level of the ``meme''---the cultural analog of the biological gene---which acts as a unit carrying cultural ideas that can be transmitted from one mind to another~\cite{lewontin1970units,shennan2002genes,hull2000taking,kuper2000if}.  While the analogy between the evolution in biology and in culture are clear (as Darwin himself has recognized), specific mechanisms at work can be very different~\cite{boyd1988culture,boyd1996culture,steward1972theory}, as in the case of meme and gene. The most important difference is that in cultural evolution there can be an arbitrary number of ``parents'' from whom a newly created work can take after since the action of inheritance takes place in the mind of the creator of new works, unlike the precisely two in sexual reproduction of organisms. This is visualized in more detail in Fig.~\ref{figure01}~(a)~and~(b). In biological evolution (Fig.~\ref{figure01}~(a)), each gene of an offspring always comes from either of the two of its parents (or rare cases of mutations or recombination error). In cultural evolution (Fig.~\ref{figure01}~(b)) on the other hand, a new work's memes can originate from an arbitrary number of ``parents'' and undergo frequent mutations.

While the study of cultural evolution is an active field~\cite{cavalli1981cultural,boyd1988culture,hewlett1986cultural,aunger2000life,mesoudi2008cultural,mcelreath2005applying,maccallum2012evolution,
pagel2009human,gray2003language,basalla1988evolution,hinde1985evolution,plotkin1981multiple,m1983diffusion}, recent developments in data and machine-learning techniques are providing even newer opportunities for understanding cultural evolution by leveraging the rich feature sets available of cultural products~\cite{aunger2002electric,shin2017chart}. In addition, the commodified nature of cultural products in the present day means that the consumer choice as selection pressure is increasing, speeding up the rate of evolution that could allows us to more easily observe the evolutionary process in a short time scale. Cultural production in today's high-risk, high-return environment of cultural production would also benefit from a deeper scientific understanding of the cultural evolutionary process.

\section{Data and Movie Memes}
We analyze the data from MovieLens, a movie recommendation service that also provides a stable benchmark dataset to researchers~\cite{grouplensurl}. The dataset contains 20 million ratings and $465,000$ tags on $27,000$ movies provided by $138,000$ users between January 1995 and March 2015. Tags are user-attached metadata that describe the movies' themes or related concepts, typically a single word or a short phrase such as ``nostalgic,'' ``artificial intelligence,'' and so forth. The $1,100$ unique tags feature numerical relevance (association) scores between $0$ and $1$ to movies, computed via machine-learning algorithms on user-contributed reviews and ratings~\cite{vig2012tag}. Tags thus contain bits of information about the movies, and each movie is composed of different combinations of such tags with varying relevance scores. Therefore we can consider each tag as a meme constituting the movies.  The parallel between gene and meme are clear from Fig.~\ref{figure01}; as an individual organism can be viewed as a set of distinct genomes, an individual movie can be viewed as a distinct set of memes. In this work we consider the $10,380$ movies that have relevance scores across all the memes. We also eliminated the tags not related to the content of movies, such as simple facts (names of directors, actors, writers, production companies), ``Oscar'', ``best of 2005'', etc. In addition, we consolidated tags that are synonyms, e.g. ``fight scenes'' and ``fighting'', using hierarchical clustering methods on their similarity in relevance scores to the movies. This grouping of tags is based on their actual relationship to the movies, rather than semantically. This process is reminiscent of identifying the so-called linked genes in cells, pairs or sets of genes located on the same chromosome inherited together in organisms. Our final data for analysis consists of $900$ unique memes. The system of movies and tags can be represented as a weighted bipartite network, mathematically a matrix of the ``meme vectors'' as follows:
\begin{eqnarray}
\textbf{M} \equiv \left\{\textit{m}_\textit{ij}\right\}
\end{eqnarray} 
where $i$ denotes a movie, $j$ denotes a tag meme and each element is between 0 and 1. The matrix is also visualized in Fig.~\ref{figure01}~(c). 

\subsection{Clusters of Movies: Genres}
We can now define the similarity between movies, for instance via the Pearson Correlation Coefficient of their meme vectors.  We can then construct a network of movies by connecting movie pairs whose similarity exceeds a certain threshold (0.8 in Fig.~\ref{figure02}).  This process is analogous to identifying one's kins via similarity in genetic makeup. The network and the communities of movies are shown in Fig.~\ref{figure02}~(a)~\cite{lambiotte2008laplacian}. A manual inspection of the detected communities shows that they correspond well to widely-accepted movie genres such as Thriller, Comedy, Romance, and so forth. They are presented along with the top-three highest-scoring memes in each genre in Fig.~\ref{figure02}~(b). These memes intuitively support the community-genre equivalence, for instance, ``Action'', ``Fight Scenes'', and ``Fast Paced'' being the top three memes for the genre that is clearly Action. Fig.~\ref{figure02}~(c) shows the movies and the networks along the movie production years on the $x$-axis. The two staple genres in terms of their size and longevity are Drama and Comedy, having been produced steadily since the very early years. It also shows that the genres increasingly cross over to one another, especially starting from the 1980s. In particular, Comedy shows a high level of interaction with Romance, clearly related to the birth of the so-called Romantic Comedy. Drama shows an increase in interaction with Horror and Action as well.  We may   label Fig.~\ref{figure02}~(c) the Phylogenetic tree~\cite{doolittle1999phylogenetic} of movies, although in culture where cross-fertilization is free to occur and thus definition of a ``species'' may be less clear-cut.

\subsection{Popular Memes and Impactful Movies}
With the movies' meme scores and production year data, we can follow the rise and fall of memes within genres, in particular the ones that enjoy a sustained status and earn a position in history. To lessen the effect of the yearly fluctuations in the number of movies produced, we use a three-year time window. Within each time window we measure the average meme score within each genre, and focus on the popular ones defined as having been among the top three high-scoring ones for three or longer consecutive time windows. The results are shown for all movies and for each genre are shown in Fig.~\ref{figure03}.  (Self-explanatory ones such as ``black and white'' and ``silent'' from the early 1900s, and ``1950s'' have been omitted from the figure.)

\begin{figure} 
\includegraphics[width=85mm]{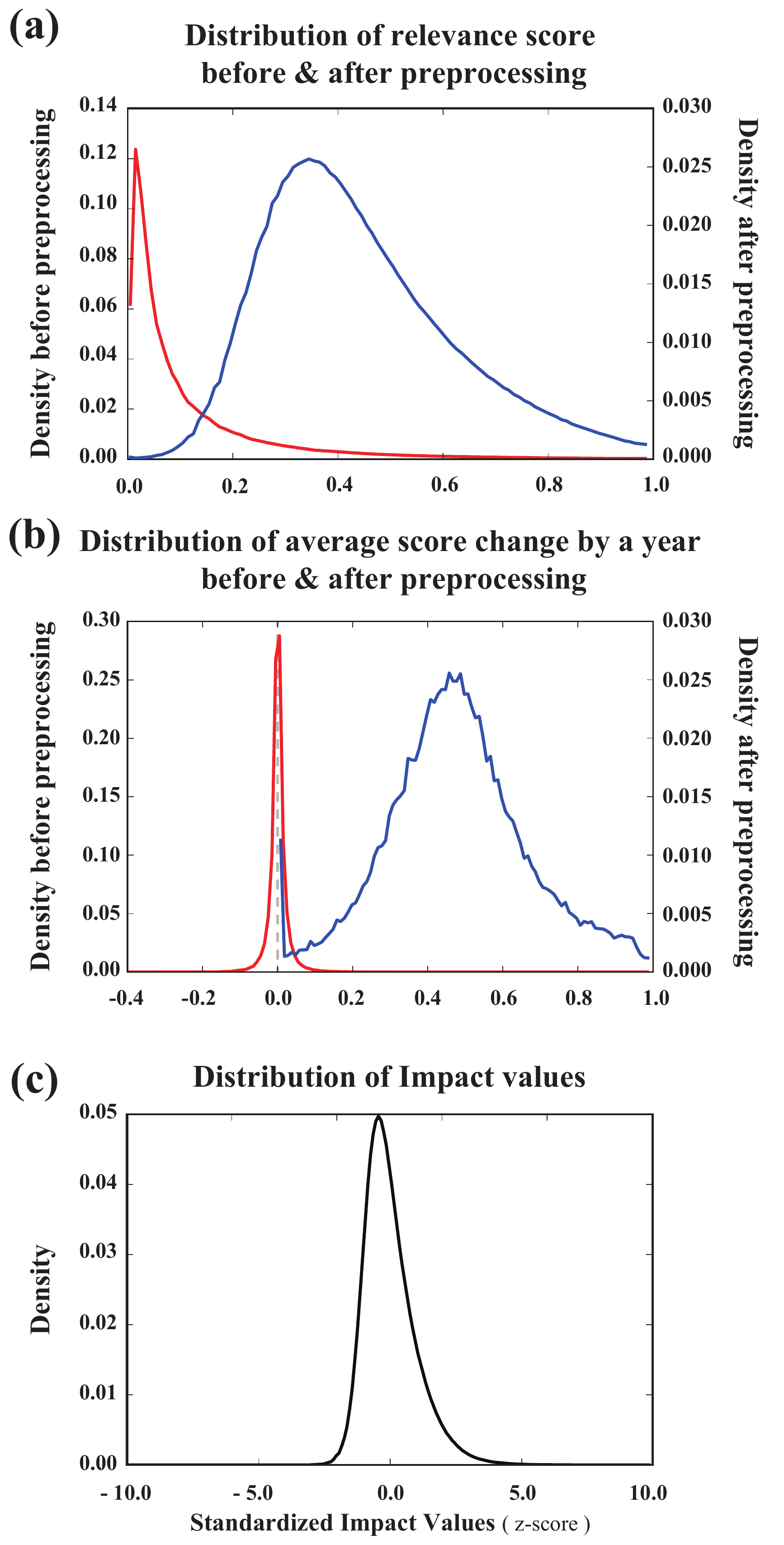} 
\caption{(a) The distribution of the meme score $m$ before (red) and after (blue) regularization. (b) The distribution of average meme score before and after any production year. The original distribution is nearly regular (red), necessitating no further processing than rescaling and translation to make it positive (blue). (c) Distribution of the impact of the movies, the product of the two variables shown in (a) and (b).} 
\label{figure05} 
\end{figure}

\begin{figure*} 
\includegraphics[width=160mm]{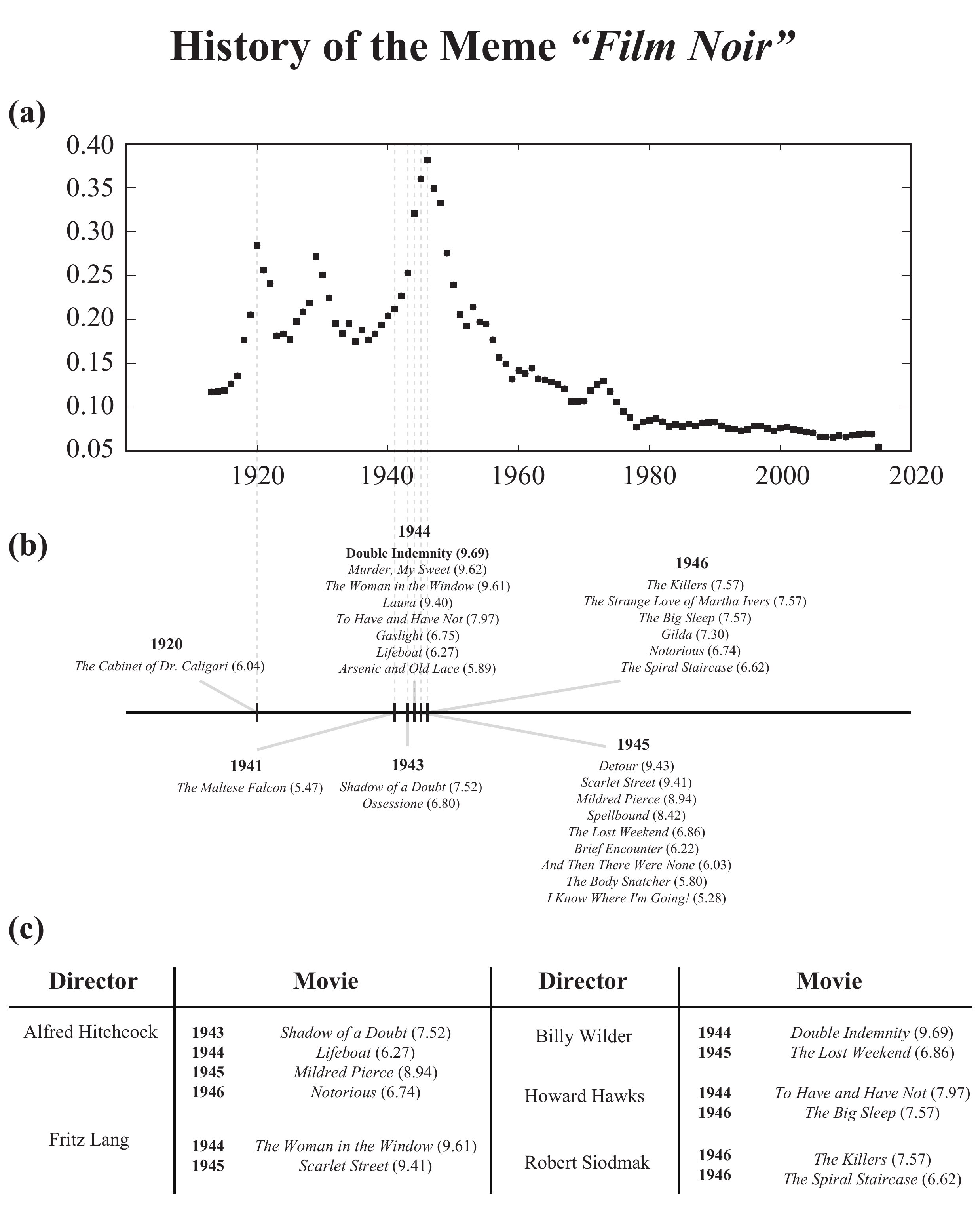} 
\caption{History of the meme ``Film Noir.'' (a) The rise and fall of the yearly average relevance score of the meme ``Film Noir''. We find three major peaks in the early years, with the highest one in the 1940s. (b) The impactful movies ($z>5$) in the meme ``Film Noir''. The values in parentheses are the $z$-scores. The most impactful movie turns out to be \emph{Double Indemnity} from 1944 by Billy Wilder, in fact considered one of the most influential in cinema studies. (c) Two directors, Alfred Hitchcock and Robert Siodmak, with the most movies in the list of impactful movies.} 
\label{figure06} 
\end{figure*}

\begin{figure*} 
\includegraphics[width=160mm]{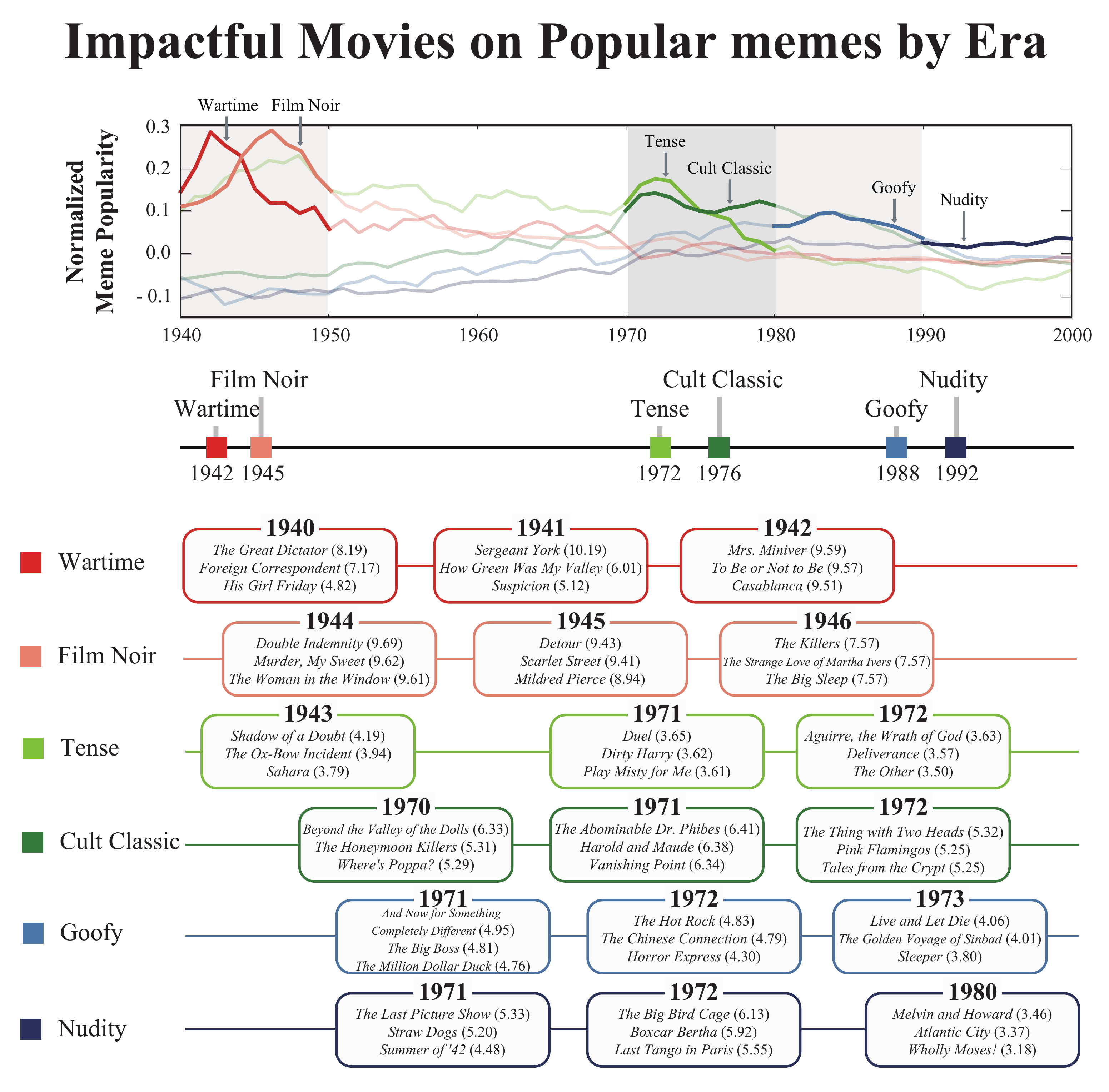} 
\caption{The six most prevalent memes by era with their years of the highest peak average relevance and the most impactful movies. The impactful movies in cinema-specific memes such as ``Wartime'' and ``Film Noir'' show higher $z$-score than those of general adjectives such as ``Tense'' and ``Goofy.'' The top one percent impactful movies of ``Wartime'' and ``Film Noir'' have higher $z$-scores than $3.27$ and $3.09$, while those of ``Tense'' and ``Goofy'' have higher $z$-scores than $2.57$ and $2.54$.} 
\label{figure07} 
\end{figure*}

\begin{figure} 
\includegraphics[width=90mm]{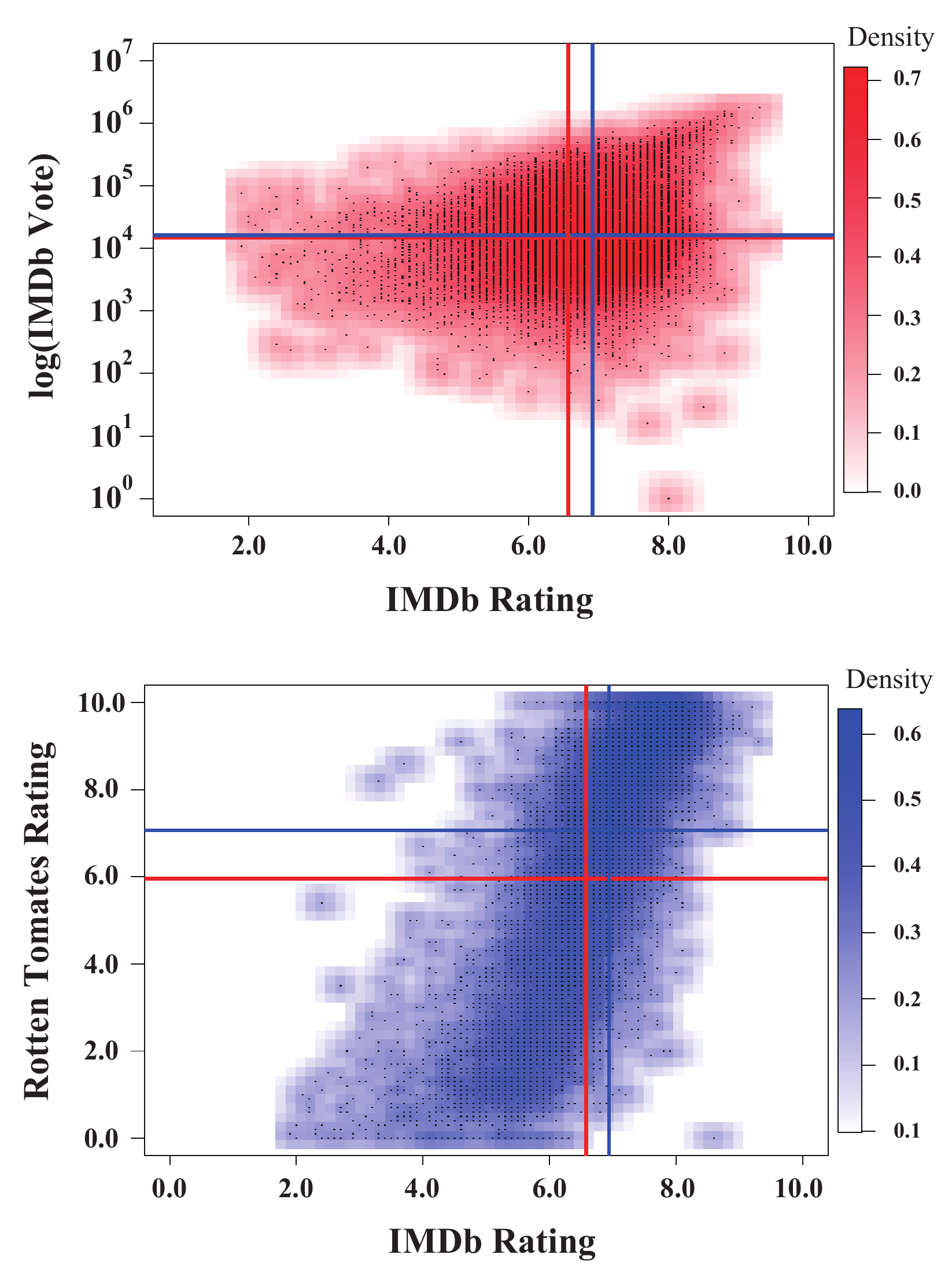} 
\caption{Impact and the evaluations by the expert and the public. (a) The relationship between IMDb Rating and IMDb Vote. The Pearson coefficient is $0.189$[errors?], showing weak correlation. The blue lines represent average values for impactful movies and the red line for non-impactful movies. IMDb Vote shows almost no correlation. (b) The relationship between IMDb rating and Rotten Tomatoes rating. The Pearson coefficient is $0.782$, showing higher correlation. The impactful movie group shows higher average rating than non-impactful movie group for both IMDb and Rotten Tomatoes rating.} 
\label{figure08} 
\end{figure}

Based on the popularity of memes throughout time, we can ask more detailed questions regarding many interesting aspects of the dynamics. For instance, in evolution the question of when a successful species appeared (the proverbial ``Eve'' for the humans) is an intriguing and enduring one. In our context, it would be asking what are the movies that acted as the harbingers of a future prominent meme. In the simplest terms, we employ the following criteria to find it: The movie itself should be highly relevant to the meme, and the movies produced after it should have a significantly higher relevance than the ones produced before it. This is illustrated in Fig.~\ref{figure04} which shows movie \textit{M} itself with a high relevance to meme $m$ that has a significantly higher score after $M$ than before it (we again consider a three-year time window).  On a technical level, to combine the two factors we first need to understand the behavior of the meme score $m$.  Fig.~\ref{figure05}~(a) shows the distribution of $m$ (red) being right-skewed, calling for regularization to make it better behaved, which in this case can be achieved by taking the logarithm $\log(m+\epsilon)$, shown in blue, with $\epsilon=0$ since $m$ are all nonzero in our data. On the other hand, the difference between mean meme scores after and before (all in a three-year window) a movie is already close to a normal distribution centered around near zero, which means we can use it as-is, though to make the product of the two terms positive, we transform the latter to take on the value in the range $(0,1)$. See Fig.~\ref{figure05}~(b). We then define the impact of a movie of meme score $m$ to be
\begin{eqnarray}
I(m) \equiv \log(m)\times (\overline{m}_{\textrm{after}}-\overline{m}_{\textrm{before}}).
\end{eqnarray}

Since we are interested in the movies with the highest impact, so we focus on those whose $z$-scores of impact $\mu\equiv(I-\overline{I})/\sigma_I$ exceed some value which set to be $5$ in our work. Fig.~\ref{figure06} shows the history of the meme ``Film Noir'' that saw particular success between the 1920s and 1940s, and its most impactful movies. The meme's average relevance score is shown in Fig.~\ref{figure06}~(a), exhibiting three major peaks with the largest one in the 1940s. Fig.~\ref{figure06}~(b) shows the impactful movies with $\mu>5$, mainly produced in the early 1940s. According to our measure the most impactful movie is Billy Wilder's \emph{Double Indemnity}, indeed considered a classic in the Film Noir genre in history. Alfred Hitchcock has the largest number of impactful movies in the genre, shown in Fig.~\ref{figure06}~(c) along with other classics. In Fig.~\ref{figure07}, we show the six most prevalent memes by era (see also Fig.~\ref{figure03}) with their years of the highest peak average relevance and the most impactful movies around them. We also find that impactful movies in cinema-specific memes such as ``Wartime'' and ``Film Noir'' tend to have in general higher $z$-score than those made of general adjectives such as ``Tense'' and ``Goofy.'' The top one percent impactful movies of ``Wartime'' and ``Film Noir'' have higher $z$-scores than $3.27$ and $3.09$, while those of ``Tense'' and ``Goofy'' have higher $z$-scores than $2.57$ and $2.54$.

\subsection{Impact and Public Evaluation}
The impact defined in our paper is based on the prevalence of memes, indirectly constructed using computational means on various types of user input. We then ask whether the impactful movies have also been received well by the audience, which could be a measure of fitness in the Darwinian sense: If a movie is not well received, the chance of passing its memes onto future movies would certainly diminish. We use the Internet Movie Database(IMDb) and Rotten Tomatoes data for this purpose. IMDb provides two types of data, IMDb Vote (the number of votes cast) and IMDb Rating (the average scores from the votes). Rotten Tomatoes, on the other hand, is an aggregation service of reviews by critics. Thus IMDb represents the general public's reception of movies, whereas Rotten Tomatoes presents the experts'. We use the log of the IMDb Vote since it is unbounded, and a few movies feature disproportionately high vote counts. Fig.~\ref{figure08} shows the relationship between these variables. IMDb Vote and IMDb Rating are only modestly correlated with Pearson correlation coefficient of $0.189\pm 9.085\times10^{-7}$, whereas IMDb Rating and Rotten Tomatoes are more strongly correlated with $0.782\pm 4.24\times10^{-7}$. More telling would be whether impactful movies differ from others with respect to these indicators. To see if this is the case, we divide the movies into two groups: Highly impactful ($\mu>5$) with regards to at least one meme (total $2,567$ out of $10,380$ movies) and the others ($7,813$ movies). We then find that the impactful movies outscore the others consistently among ratings, averaging $6.92$ (IMDb Rating) and $7.07$ (Rotten Tomatoes), respectively, versus $6.56$ and $5.96$ ($p=.000$). The IMDb Vote, however, turns out to be nondiscriminatory for the two groups ($4.21$ and $4.18$, $p=.054$). This tells us that the sheer interest in a movie is likely unrelated to the eventual ratings given by the viewership, and has more to do with hype or marketing.

\section{Conclusion}  
Here we have proposed studying the evolutionary dynamics of cultural systems at the meme level. The meme, the biological analog of the gene, is a human-generated tag that had variable relevance to different cultural products. We were able to construct the network of movies based on meme-level similarity, discovering the community structure that corresponded well with common genre designations. We also found that cross-genre connections increased over time, indicating rising complexity in meme compositions. We observed clear evidence of the memes rising and falling in popularity, and identified the impactful movies that herald the success of a meme. Furthermore, we examined the viewership's reception of those movies and found positive correlations, meaning that for a movie to impact the future and give rise to a new successful meme it must appeal to the consumer base as well, not merely be different from its predecessors.

The general nature of our framework and findings suggest it can be applied in many diverse cultural areas where similar data are available, which is becoming increasingly the case thanks to massive consumer participation and advances in data collection.  In an era where cultural products are produced systematically and presented to an ever-widening pool of consumers, the works of this kind to understand how culture evolves due to market pressure has the potential to lead to valuable insights on how culture, technology, and the society interact.

\bibliography{movielens_bib}  
 
\end{document}